\documentclass[graybox,vecphys]{svmult}

\usepackage{mathptmx}       
\usepackage{helvet}         
\usepackage{courier}        
\usepackage{type1cm}        
\usepackage{makeidx}         
\usepackage{graphicx}        
\usepackage{multicol}        
\usepackage[bottom]{footmisc}
\usepackage{amsmath,amssymb}

\renewcommand{\r}[1]{\mathrm{#1}}
\newcommand{\mat}[1]{\underline{\underline{#1}}}
\newcommand{\vv}[1]{\underline{#1}}

\makeindex             

\begin{document}

\title*{Global parameter identification of stochastic reaction networks from single trajectories}
\titlerunning{Global stochastic parameter identification}
\author{Christian L.~M\"uller$^\dag$, Rajesh Ramaswamy$^\dag$, and Ivo F.~Sbalzarini}
\authorrunning{C.~L.~M\"uller, R.~Ramaswamy, and I.~F.~Sbalzarini}
\institute{Christian L.~M\"uller, Rajesh Ramaswamy, and Ivo F.~Sbalzarini: \at Institute of Theoretical Computer Science and Swiss Institute of Bioinformatics, ETH Zurich, CH--8092 Zurich, Switzerland, \email{christian.mueller@inf.ethz.ch, rajeshr@ethz.ch, ivos@ethz.ch}. \\
$^\dag$ These authors contributed equally to this work.}
%
%
\maketitle

\abstract*{We consider the problem of inferring the unknown parameters of a stochastic biochemical network model from a single measured time-course of the concentration of some of the involved species. Such measurements are available, e.g., from live-cell fluorescence microscopy in image-based systems biology. In addition, fluctuation time-courses from, e.g., fluorescence correlation spectroscopy provide additional information about the system dynamics that can be used to more robustly infer parameters than when considering only mean concentrations. Estimating model parameters from a single experimental trajectory enables single-cell measurements and quantification of cell--cell variability. We propose a novel combination of an adaptive Monte Carlo sampler, called Gaussian Adaptation, and efficient exact stochastic simulation algorithms that allows parameter identification from single stochastic trajectories. We benchmark the proposed method on a linear and a non-linear reaction network at steady state and during transient phases. In addition, we demonstrate that the present method also provides an ellipsoidal volume estimate of the viable part of parameter space and is able to estimate the physical volume of the compartment in which the observed reactions take place.}

\abstract{We consider the problem of inferring the unknown parameters of a stochastic biochemical network model from a single measured time-course of the concentration of some of the involved species. Such measurements are available, e.g., from live-cell fluorescence microscopy in image-based systems biology. In addition, fluctuation time-courses from, e.g., fluorescence correlation spectroscopy provide additional information about the system dynamics that can be used to more robustly infer parameters than when considering only mean concentrations. Estimating model parameters from a single experimental trajectory enables single-cell measurements and quantification of cell--cell variability. We propose a novel combination of an adaptive Monte Carlo sampler, called Gaussian Adaptation, and efficient exact stochastic simulation algorithms that allows parameter identification from single stochastic trajectories. We benchmark the proposed method on a linear and a non-linear reaction network at steady state and during transient phases. In addition, we demonstrate that the present method also provides an ellipsoidal volume estimate of the viable part of parameter space and is able to estimate the physical volume of the compartment in which the observed reactions take place.}
\section{Introduction}\label{sec:intro}

Systems biology implies a holistic research paradigm, complementing the reductionist approach to biological organization \cite{Kitano:2002b,Kitano:2002a}. This frequently has the goal of mechanistically understanding the function of biological entities and processes in interaction with the other entities and processes they are linked to or communicate with. A formalism to express these links and connections is provided by network models of biological processes \cite{Barabasi:2004,Albert:2005}. Using concepts from graph theory \cite{Mason:2007} and dynamic systems theory \cite{Wolkenhauer:2001}, the organization, dynamics, and plasticity of these networks can then be studied.

Systems biology models of molecular reaction networks contain a number of parameters. These are the rate constants of the involved reactions and, if spatiotemporal processes are considered, the transport rates, e.g.~diffusion constants, of the chemical species. In order for the models to be predictive, these parameters need to be inferred. The process of inferring them from experimental data is called {\em parameter identification}. If in addition also the network structure is to be inferred from data, the problem is called {\em systems identification}. Here, we consider the problem of identifying the parameters of a biochemical reaction network from a single, noisy measurement of the concentration time-course of some of the involved species. While this time series can be long, ensemble replicas are not possible, either because the measurements are destructive or one is interested in variations between different specimens or cells. This is particularly important in {\em molecular systems biology}, where cell--cell variations are of interest or large numbers of experimental replica are otherwise not feasible \cite{Zechner:2011a}. 

This problem is particularly challenging and traditional genomic and proteomic techniques do not provide single-cell resolution. Moreover, in individual cells the molecules and chemical reactions can only be observed indirectly. Frequently, fluorescence microscopy is used to observe biochemical processes in single cells. Fluorescently tagging some of the species in the network of interest allows measuring the spatiotemporal evolution of their concentrations from video microscopy and fluorescence photometry. In addition, fluorescence correlation spectroscopy (FCS) allows measuring fluctuation time-courses of molecule numbers \cite{Lakowicz:2006}.

Using only a single trajectory of the mean concentrations would hardly allow identification of network parameters. There could be several combinations of network parameters that lead to the same mean dynamics. A stochastic network model, however, additionally provides information about the fluctuations of the molecular abundances. The hope is that there is then only a small region of parameter space that produces the correct behavior of the mean {\em and} the correct spectrum of fluctuations \cite{Munsky:2009}. Experimentally, fluctuation spectra can be measured at single-cell resolution using FCS. 

The stochastic behavior of biochemical reaction networks can be due to low copy numbers of the reacting molecules \cite{Ramaswamy:2011a,Grima:2009}. In addition, biochemical networks may exhibit stochasticity due to extrinsic noise. This can persist even at the continuum scale, leading to continuous--stochastic models. Extrinsic noise can, e.g., arise from environmental variations or variations in how the reactants are delivered into the system. Also measurement uncertainties can be accounted for in the model as extrinsic noise, modeling our inability to precisely quantify the experimental observables. 

We model stochastic chemical kinetics using the chemical master equation (CME). Using a CME forward model in biological parameter identification amounts to tracking the evolution of a probability distribution, rather than just a single function. This prohibits predicting the state of the system and only allows statements about the probability for the system to be in a certain state, hence requiring sampling-based parameter identification methods. 
In the stochastic--discrete context, a number of different approaches have been suggested. Boys \textit{et al.} proposed a fully Bayesian approach for parameter estimation using an explicit likelihood for data/model comparison and a Markov Chain Monte Carlo (MCMC) scheme for sampling \cite{Boys:2008}. Zechner \textit{et al.} developed a recursive Bayesian estimation technique \cite{Zechner:2011,Zechner:2011a} to cope with cell--cell variability in experimental ensembles. Toni and co-workers used an approximate Bayesian computation (ABC) ansatz, as introduced by Marjoram and co-workers \cite{Marjoram:2003}, that does not require an explicit likelihood \cite{Toni:2009}. Instead, sampling is done in a sequential Monte Carlo (or particle filter) framework. Reinker \textit{et al.} used a hidden Markov model where the hidden states are the actual molecule abundances and state transitions model chemical reactions \cite{Reinker:2006}. Inspired by Prediction Error Methods \cite{Ljung:2002}, Cinquemani \textit{et al.} identified the parameters of a hybrid deterministic--stochastic model of gene expression from multiple experimental time courses \cite{Cinquemani:2008}. Randomized optimization algorithms have been used, e.g., by Koutroumpas \textit{et al.} who applied a Genetic Algorithm to a hybrid deterministic--stochastic network model \cite{Koutroumpas:2008}. More recently, Poovathingal and Gunawan used another global optimization heuristic, the Differential Evolution algorithm \cite{Poovathingal:2010}. A variational approach for stochastic two-state systems is proposed by Stock and co-workers based on Maximum Caliber \cite{Stock:2008}, an extension of Jaynes' Maximum Entropy principle \cite{Jaynes:1957} to non-equilibrium systems. If estimates are to be made based on a single trajectory, the stochasticity of the measurements and of the model leads to very noisy similarity measures, requiring optimization and sampling schemes that are robust against noise in the data.

Here, we propose a novel combination of exact stochastic simulations for a CME forward model and an adaptive Monte Carlo sampling technique, called Gaussian Adaptation, to address the single-trajectory parameter estimation problem for monostable stochastic biochemical reaction networks. Evaluations of the CME model are done using exact partial-propensity stochastic simulation algorithms \cite{Ramaswamy:2009}. Parameter optimization uses Gaussian Adaptation. The method iteratively samples model parameters from a multivariate normal distribution and evaluates a suitable objective function that measures the distance between the dynamics of the forward model output and the experimental measurements. In addition to estimates of the kinetic parameters in the network, the present method also provides an ellipsoidal volume estimate of the viable part of parameter space and is able to estimate the physical volume of the intra-cellular compartment in which the reactions take place.

We assume that quantitative experimental time series of either a transient or the steady state of the concentrations of some of the molecular species in the network are available. This can, for example, be obtained from single-cell fluorescence microscopy by translating fluorescence intensities to estimated chemical concentrations. Accurate methods that account for the microscope's point-spread function and the camera noise model are available to this end \cite{Helmuth:2009,Helmuth:2009a,Cardinale:2009}. Additionally, FCS spectra can be analyzed in order to quantify molecule populations, their intrinsic fluctuations, and lifetimes \cite{Lakowicz:2006,Qian:2004,Ramaswamy:2011a}. The present approach requires only a {\em single} stochastic trajectory from each cell. Since the forward model is stochastic and only a single experimental trajectory is used, the objective function needs to robustly measure closeness between the experimental and the simulated trajectories. We review previously considered measures and present a new distance function in Sec.~\ref{sec:costfunction}. First, however, we set out the formal stochastic framework and problem description below. We then describe Gaussian Adaptation and its applicability to the current estimation task. The evaluation of the forward model is outlined in Sec.~\ref{sec:forwardmodel}. We consider a cyclic linear chain as well as a non-linear colloidal aggregation model as benchmark test cases in Sec.~\ref{sec:results} and conclude in Sec.~\ref{sec:discuss}.

\section{Background and problem statement}

We consider a network model of a biochemical system given by $M$ coupled chemical reactions 
\begin{equation}
\sum_{i=1}^N \nu^-_{i,j}\r{S}_{i} \xrightarrow{\quad k_j\quad} \sum_{i=1}^N \nu^+_{i,j}\r{S}_{i} \quad \forall j=1,\ldots, M  
\label{eq:system}
\end{equation}
between $N$ species, where ${\mat{\nu}^-} = [ \nu^-_{i,j} ]$ and ${\mat{\nu}^+} = [ \nu^+_{i,j} ]$ are the stoichiometry matrices of the reactants and products, respectively, and $\r{S}_{i}$ is the $i^{\r{th}}$ species in the reaction network. Let $n_{i}$ be the population (molecular copy number) of species $\r{S}_{i}$. 
The reactions occur in a physical volume $\Omega$ and the macroscopic reaction rate of reaction $j$ is $k_j$. 
This defines a dynamic system with state ${\vv{n}}(t) = [ n_i(t) ]$ and $M+1$ parameters ${\vv{\theta}} = [ k_1,\ldots, k_M, \Omega ]$.  

The state of such a system can be interpreted as a realization of a random variable ${\vv{n}}(t)$ that changes over time $t$. All one can know about the system is the probability for it to be in a certain state at a certain time $t_j$ given the system's state history, hence 
\begin{eqnarray}
&& P(\vv{n}(t_j) \,|\,{\vv{n}}(t_{j-1}),\ldots,{\vv{n}}(t_{1}),{\vv{n}}(t_{0}))\,\mathrm{d}^N\!{n} \notag \\
&& \quad = \mathrm{Prob}\{{\vv{n}}(t_j) \in [{\vv{n}}(t_j),\, {\vv{n}}(t_j) + \mathrm{d}{\vv{n}})\,|\, {\vv{n}}(t_i), \,\,i=0,\ldots,j-1\,\} \, .
\end{eqnarray}

A frequently made model assumption, substantiated by physical reasoning, is that the probability of the current state depends solely on the previous state, i.e.,
\begin{eqnarray}
&& P(\vv{n}(t_j)  \,|\,{\vv{n}}(t_{j-1}),\ldots,{\vv{n}}(t_{1}),{\vv{n}}(t_{0})) = P(\vv{n}(t_j) \,|\,{\vv{n}}(t_{j-1})) \, .
\end{eqnarray}
The system is then modeled as a first-order Markov chain where the state ${\vv{n}}$ evolves as 
\begin{equation}
{\vv{n}}(t+\Delta t) = {\vv{n}}(t) + {\vv{\Xi}}(\Delta t;\,\vv{n}, t)
\end{equation} 
This is the {\em equation of motion} of the system. If ${\vv{n}}$ is real-valued, it defines a continuous--stochastic model in the form of a continuous-state Markov chain. Discrete ${\vv{n}}$, as is the case in chemical kinetics, amount to discrete--stochastic models expressed as discrete-state Markov chains. The Markov propagator ${\vv{\Xi}}$ is itself a random variable, distributed with probability distribution $\Pi({\vv{\xi}} \,|\,\Delta t;\,{\vv{n}},t) = P({\vv{n}}+{\vv{\xi}},t+\Delta t\,|\, {\vv{n}},t)$ for the state change ${\vv{\xi}}$. For continuous-state Markov chains, $\Pi$ is a continuous probability density function (PDF), for discrete-state Markov chains a discrete probability distribution. If $\Pi({\vv{\xi}})=\delta({\vv{\xi}}-{\vv{\xi}}_0)$, with $\delta$ the Dirac delta distribution, then the system's state evolution becomes deterministic with predictable discrete or continuous increments ${\vv{\xi}}_0$. Deterministic models can hence be interpreted as a limit case of stochastic models \cite{Kurtz:1972}. 

In chemical kinetics, the probability distribution $\Pi$ of the Markov propagator is a linear combination of Poisson distributions with weights given by the reaction stoichiometry. This leads to the equation of motion for the population ${\vv{n}}$ given by
\begin{equation}
{\vv{n}}(t+\Delta t) = {\vv{n}}(t) + ({\mat{\nu}^+}-{\mat{\nu}^-})\left[\begin{array}{c} \psi_1 \\ \ldots \\ \psi_M \end{array}\right] \, ,
\label{eq:CME-eq-of-motion}
\end{equation}
where $\psi_i \sim \mathcal{P}(a_i({\vv{n}}(t))\Delta t)$ is a random variable from the Poisson distribution with rate $\lambda=a_i({\vv{n}}(t))\Delta t$. The second term on the right-hand side of Eq.~\ref{eq:CME-eq-of-motion} follows a probability distribution $\Pi({\vv{\xi}}\,|\,\Delta t; {\vv{n}},t)$ whose explicit form is analytically intractable in the general case. The rates $a_j$, $j=1,\ldots M$, are called the {\em reaction propensities} and are defined as 
\begin{equation}
a_j = \prod_{i=1}^{N}\left(\begin{array}{c} n_i \\ \nu^{-}_{i,j} \end{array}\right) \frac{k_j}{\Omega ^{1+\sum_{i'=1}^N \nu^-_{i',j}}} \,.
\label{eq:propensity}
\end{equation}
They depend on the macroscopic reaction rates and the reaction volume and can be interpreted as the probability rates of the respective reactions. Advancing Eq.~\ref{eq:CME-eq-of-motion} with a $\Delta t$ such that more than one reaction event happens per time step yields an approximate simulation of the  biochemical network as done in approximate stochastic simulation algorithms~\cite{Gillespie:2001, Auger:2006a}.

An alternative approach consists in considering the evolution of the state probability distribution $P({\vv{n}},t\,|\,{\vv{n}_0}, t_0)$ of the Markov chain described by Eq.~\ref{eq:CME-eq-of-motion}, hence:
\begin{equation}
\frac{\partial P}{\partial t} = \sum_{j=1}^M \left( \prod_{i=1}^N \r{E}_i^{\nu^-_{i,j}}\r{E}_j^{-\nu^+_{i,j}}-1\right) a_j({\vv{n}}(t)) P({\vv{n}},t)
\label{eq:CME}
\end{equation}
with the {\em step operator} $\r{E}_i^p f({\vv{n}}) = f({\vv{n}}+p \vv{\hat{i}}\,)$ for any function $f$ where $\vv{\hat{i}}$ is the $N$-dimensional unit vector along the $i^\r{th}$ dimension. This equation is called the {\em chemical master equation} (CME). Directly solving it for $P$ is analytically intractable, but trajectories of the Markov chain governed by the unknown state probability $P$ can be sampled using exact stochastic simulation algorithms (SSA)~\cite{Gillespie:1992}. Exact SSAs are exact in the sense that they sample Markov chain realizations from the exact solution $P$ of the CME, without ever explicitly computing this solution. Since SSAs are Monte Carlo algorithms, however, a sampling error remains.

Assuming that the population ${\vv{n}}$ increases with the volume $\Omega$, ${\vv{n}}$ can be approximated as a continuous random variable in the limit of large volumes, and Eq.~\ref{eq:CME-eq-of-motion} becomes
\begin{equation}
{\vv{n}}(t+\Delta t) = {\vv{n}}(t) + ({\mat{\nu}^+}-{\mat{\nu}^-})\left[\begin{array}{c} \eta_1 \\ \ldots \\ \eta_M \end{array}\right] \, ,
\label{eq:CLE-eq-of-motion}
\end{equation}
where $\eta_i \sim \mathcal{N}(a_i({\vv{n}}(t))\Delta t, \, a_i({\vv{n}}(t))\Delta t)$ are normally distributed random variables. The second term on the right-hand side of this equation is a random variable that is distributed according to the corresponding Markov propagator $\Pi({\vv{\xi}}\,|\,\Delta t; {\vv{n}},t)$, which is a Gaussian. Equation \ref{eq:CLE-eq-of-motion} is called the {\em chemical Langevin equation} with $\Pi$ given by
\begin{equation}
\Pi({\vv{\xi}}\,|\,\Delta t; {\vv{n}}, t) = (2\pi)^{-N/2} |{\mat{\Sigma}}|^{-1/2} \r{e}^{-\frac{1}{2} ({\vv{\xi}} - {\vv{\mu}})^\r{T} {\mat{\Sigma}}^{-1} ({\vv{\xi}} - {\vv{\mu}})} \, , 
\end{equation} 
where 
\[
{\vv{\mu}} = \Delta t\,({\mat{\nu}}^+ - {\mat{\nu}}^-)\left[\begin{array}{c} a_1({\vv{n}}(t)) \\ \ldots \\ a_M({\vv{n}}(t)) \end{array}\right] \text{ and\,\, } {\mat{\Sigma}} = \Delta t\,({\mat{\nu}}^+ - {\mat{\nu}}^-)\,\r{diag}\,({\vv{n}(t)}) ({\mat{\nu}}^+ - {\mat{\nu}}^-)^{\r{T}} .
\]
The corresponding equation for the evolution of the state PDF is the non-linear Fokker-Planck equation, given by
\begin{equation}
\frac{\partial P}{\partial t} = \vv{\nabla}^\r{T} \, \left(\frac{1}{2}{\mat{D}}\,\vv{\nabla}- {\vv{F}}\right)P({\vv{n}},t) \, , 
\end{equation}
where
\begin{eqnarray}
\vv{\nabla}^\r{T} = \left[\frac{\partial}{\partial n_1}, \ldots, \frac{\partial}{\partial n_N}\right],
\end{eqnarray}
\begin{eqnarray}
F_{i} &=& \lim_{\Delta t \to 0} \frac{1}{\Delta t} \int_{-\infty}^{+\infty} \D \xi_i\,\xi_i\, \Pi({\vv{\xi}}\,|\,\Delta t ; {\vv{n}},t) \ , 
\end{eqnarray}
and
\begin{eqnarray}
D_{ij} &=& \lim_{\Delta t \to 0} \frac{1}{\Delta t} \int_{-\infty}^{+\infty} \int_{-\infty}^{+\infty} \D \xi_i \D \xi_j\,\xi_i\,\xi_j\, \Pi({\vv{\xi}}\,|\,\Delta t ; {\vv{n}},t) \, .
\end{eqnarray}

At much larger $\Omega$, when the population ${\vv{n}}$ is on the order of Avogadro's number, Eq.~\ref{eq:CLE-eq-of-motion} can be further approximated as 
\begin{equation}
{\vv{n}}(t+\Delta t) = {\vv{n}}(t) + ({\mat{\nu}^+}-{\mat{\nu}^-})\left[\begin{array}{c} \phi_1({\vv{n}}(t))\Delta t \\ \ldots \\ \phi_M({\vv{n}}(t))\Delta t \end{array}\right] \, ,   
\label{eq:largeOmega}
\end{equation} 
where $\phi_j (\vv{n})= k_j \Omega ^{1-\sum_{i'=1}^N \nu^-_{i'\!,j}} \prod_{i=1}^{N} n_i^{\nu^-_{i,j}} (\nu^-_{i,j}!)^{-1}$. Note that the second term on the right-hand side of this equation is a random variable whose probability distribution is the Dirac delta
\begin{equation}
\Pi({\vv{\xi}}\,|\,\Delta t; {\vv{n}}, t) = \delta\left({\vv{\xi}} - ({\mat{\nu}^+}-{\mat{\nu}^-})\left[\begin{array}{c} \phi_1({\vv{n}}(t))\Delta t \\ \ldots \\ \phi_M({\vv{n}}(t))\Delta t \end{array}\right]\right) \, .
\end{equation}
Equation \ref{eq:largeOmega} hence is a deterministic equation of motion. In the limit $\Delta t \rightarrow 0$ this equation can be written as the ordinary differential equation
\begin{equation}
\frac{\D {\vv{x}}}{\D t} = ({\mat{\nu}^+}-{\mat{\nu}^-})\left[\begin{array}{c} \phi_1({\vv{x}}(t)) \\ \ldots \\ \phi_M({\vv{x}}(t)) \end{array}\right]
\label{eq:rre}
\end{equation} 
for the {\em concentration} ${\vv{x}} = {\vv{n}}\Omega^{-1}$. 
This is the classical {\em reaction rate equation} for the system in Eq.~\ref{eq:system}. 

By choosing the appropriate probability distribution $\Pi$ of the Markov propagator, one can model reaction networks in different regimes: small population ${\vv{n}}$ (small $\Omega$) using SSA over Eq.~\ref{eq:CME}, intermediate population (intermediate $\Omega$) using Eq.~\ref{eq:CLE-eq-of-motion}, and large population (large $\Omega$) using Eq.~\ref{eq:rre}. The complete model definition therefore is $\mathcal{M}({\vv{\theta}}) = \{{\mat{\nu}^-},{\mat{\nu}^+},\Pi$\}. 

The problem considered here can then be formalized as follows: Given a forward model $\mathcal{M}({\vv{\theta}})$ and a single noisy trajectory of the population of the  chemical species $\hat{{\vv{n}}}(t_0 + (q-1)\Delta t_\r{exp})$ at $K$ discrete time points $t = t_0 + (q-1) \Delta t_\r{exp}$, $q=1,\ldots , K$, we wish to infer ${\vv{\theta}} = [ k_1,\ldots, k_M, \Omega ]$. The time between two consecutive measurements $\Delta t_\r{exp}$ and the number of measurements $K$ are given by the experimental technique used. As a forward model we use the full CME as given in Eq.~\ref{eq:CME} and sample trajectories from it using the partial-propensity formulation of Gillespie's exact SSA as described in Sec.~\ref{sec:forwardmodel}.

\section{Gaussian Adaptation for global parameter optimization, approximate Bayesian computation, and volume estimation} \label{sec:GaA}

Gaussian Adaptation (GaA), introduced in the late 1960's by Gregor Kjellstr\"om \cite{Kjellstrom:1969,Kjellstrom:1981}, is a Monte Carlo technique that has originally been developed to solve design-centering and optimization problems in analog electric circuit design. Design centering solves the problem of determining the nominal values (resistances, capacitances, etc.) of the components of a circuit such that the circuit output is within specified design bounds and is maximally robust against random variations in the circuit components with respect to a suitable criterion or objective function. This problem is a superset of  general optimization, where one is interested in finding a parameter vector that minimizes (or maximizes) the objective function without any additional robustness criterion. GaA has been specifically designed for scenarios where the objective function $f(\vv{\theta})$ is only available in a black-box (or oracle) model that is defined on a real-valued domain $\mathcal{A} \subseteq \mathbb{R}^n$ and returns real-valued output $\mathbb{R}$. The black-box model assumes that gradients or higher-order derivatives of the objective function may not exist or may not be available, hence including the class of discontinuous and noisy functions. The specific black-box function used here is presented in Sec.~\ref{sec:costfunction}. 

The principle idea behind GaA is the following: Starting from a user-defined point in parameter space, GaA explores the space by iteratively sampling single parameter vectors from a multivariate Gaussian distribution $\mathcal{N}(\vv{m},\mat{\Sigma})$ whose mean $\vv{m} \in  \mathbb{R}^n$ and covariance matrix $\mat{\Sigma}\in  \mathbb{R}^{n \times n}$ are dynamically adapted based on the information from previously accepted samples. The acceptance criterion depends on the specific mode of operation, i.e., whether GaA is used as an optimizer or as a sampler \cite{Muller:2010a,Muller:2010b}. Adaptation is performed such as to maximize the entropy of the search distribution under the constraint that acceptable search points are found with a predefined, fixed hitting (success) probability  $p<1$ \cite{Kjellstrom:1981}. Using the definition of the entropy of a multivariate Gaussian distribution $\mathcal{H}(\mathcal{N}) = \log \left( \sqrt{(2 \pi \E)^n \det(\mat{\Sigma})} \right)$ shows that this is equivalent to maximizing the determinant of the covariance matrix $\mat{\Sigma}$. GaA thus follows Jaynes' Maximum Entropy principle \cite{Jaynes:1957}. 

GaA starts by setting the mean $\vv{m}^{(0)}$ of the multivariate Gaussian to an initial acceptable point $\vv{\theta}^{(0)}$ and the Cholesky factor $\mat{Q}^{(0)}$ of the  covariance matrix to the identity matrix $\mat{I}$. At each iteration $g>0$, the covariance $\mat{\Sigma}^{(g)}$ is decomposed as: $\mat{\Sigma}^{(g)} = \left(r\cdot\mat{Q}^{(g)}\right)\left(r\cdot\mat{Q}^{(g)}\right)^{\!\r{T}} = r^2 \left(\mat{Q}^{(g)}\right)\left(\mat{Q}^{(g)}\right)^{\!\r{T}},
$
where $r$ is the scalar step size that controls the scale of the search. The matrix $\mat{Q}^{(g)}$ is the normalized square root of $\mat{\Sigma}^{(g)}$, found by eigen- or Cholesky decomposition of $\mat{\Sigma}^{(g)}$. The candidate parameter vector in iteration \mbox{$g+1$} is sampled from a multivariate Gaussian according to $\vv{\theta}^{(g+1)} = \vv{m}^{(g)} + r^{(g)}  \mat{Q}^{(g)} \vv{\eta}^{(g)}$, where $\vv{\eta}^{(g)} \sim \mathcal{N}(\vv{0},  \mat{I})$. The parameter vector is then evaluated by the objective function $f(\vv{\theta}^{(g+1)})$. 

{\em Only if} the parameter vector is accepted, the following adaptation rules are applied: The step size $r$ is increased as $r^{(g+1)} = f_\text{e} \cdot r^{(g)}$, where $f_\text{e}>1$ is termed the {\em expansion factor}. The mean of the proposal distribution is updated as
\begin{equation}
\vv{m}^{(g+1)} = \left(1-\frac{1}{N_\text{m}}\right)\vv{m}^{(g)} +\frac{1}{N_\text{m}} \vv{\theta}^{(g+1)} \, .
\label{eq:GaA_m}
\end{equation}
$N_\text{m}$ is a weighting factor that controls the learning rate of the method. The successful search {\em direction} $\vv{d}^{(g+1)}=\left(\vv{\theta}^{(g+1)}-\vv{m}^{(g)}\right)$ is used to perform a rank-one update of the covariance matrix: $\mat{\Sigma}^{(g+1)} = \left(1-\frac{1}{N_{\text{C}}}\right)\mat{\Sigma}^{(g)}+ \frac{1}{N_{\text{C}}}\vv{d}^{(g+1)} \vv{d}^{(g+1) \, T}$. $N_\text{C}$ weights the influence of the accepted parameter vector on the covariance matrix. In order to decouple the volume of the covariance (controlled by $r^{(g+1)} $) from its orientation, $\mat{Q}^{(g+1)} $ is normalized such that $\det(\mat{Q}^{(g+1)})=1$. 

In case $\vv{\theta}^{(g+1)}$ is not accepted at the current iteration, only the step size is adapted as
$ r^{(g+1)} = f_\text{c} \cdot r^{(g)}$, where $f_\text{c}<1$ is the {\em contraction factor}. 

The behavior of GaA is controlled by several strategy parameters. Kjellstr\"om analyzed the information-theoretic optimality of the acceptance probability $p$ for GaA in general regions \cite{Kjellstrom:1981}. He concluded that the efficiency $E$ of the process and $p$ are related as $E \propto -p \log{p}$, leading to an optimal $p=\frac{1}{\r{e}} \approx 0.3679$, where $\r{e}$ is Euler's number. A proof is provided in \cite{Kjellstrom:1991}. Maintaining this optimal hitting probability corresponds to leaving the volume of the distribution, measured by $\det(\mat{\Sigma})$, constant under stationary conditions. Since $\det(\mat{\Sigma})=r^{2n}\det(\mat{Q}\, \mat{Q}^\mathrm{T})$, the expansion and contraction factors $f_\text{e}$ and $f_\text{c}$ expand or contract the volume by a factor of $f_\text{e}^{2n}$ and $f_\text{c}^{2n}$, respectively. After $S$ accepted and $F$ rejected samples, a necessary condition for constant volume thus is: $\prod_{i=1}^{S} (f_\text{e})^{2n} \prod_{i=1}^{F} (f_\text{c})^{2n} =1$. Using $p=\frac{S}{S+F}$, and introducing a small $\beta>0$, the choice $f_\text{e} = 1+ \beta(1-p)$ and $f_\text{c} = 1- \beta p$ satisfies the constant-volume condition to first order. The scalar rate $\beta$ is coupled to $N_\text{C}$. $N_\text{C}$ influences the update of  $\mat{\Sigma} \in \mathbb{R}^{n \times n}$, which contains $n^2$ entries.  Hence, $N_\text{C}$ should be related to $n^2$. We suggested using $N_\text{C}=(n+1)^2 / \log(n+1)$ as a standard value, and coupling $\beta=\frac{1}{N_\text{C}}$ \cite{Muller:2010}. A similar reasoning is also applied to $N_\text{m}$. Since $N_\text{m}$ influences the update of $\mat{m} \in \mathbb{R}^{n}$, it is reasonable to set $N_\text{m} \propto n$. We propose $N_\text{m}=\r{e}n$ as a standard value. 

Depending on the specific acceptance rule used, GaA can be turned into a global optimizer \cite{Muller:2010}, an adaptive MCMC sampler \cite{Muller:2010a,Muller:2010b}, or a volume estimation method \cite{Muller:2011c}, as described next.

\subsection{GaA for global black-box optimization}
In a minimization scenario, GaA uses an adaptive-threshold acceptance mechanism. Given an initial scalar cutoff threshold $c_\text{T}^{(0)}$, we accept a parameter vector $\vv{\theta}^{(g+1)}$ at iteration \mbox{$g+1$} if $f(\vv{\theta}^{(g+1)})< c_\text{T}^{(g)}$. Upon acceptance, the threshold $c_\text{T}$ is lowered as $c_\text{T}^{(g+1)} = \left(1-\frac{1}{N_\text{T}}\right)c_\text{T}^{(g)} +\frac{1}{N_\text{T}} f(\vv{\theta}^{(g+1)})$, where $N_\text{T}$ controls the weighting between the old threshold and the objective-function value of the \emph{accepted} sample. This sample-dependent threshold update renders the algorithm invariant to linear transformations of the objective function. The standard strategy parameter value is $N_\text{T}=\r{e}n$ \cite{Muller:2010a}. We refer to \cite{Muller:2010a} for further information about convergence criteria and constraint handling techniques in GaA.  

\subsection{GaA for approximate Bayesian computation and viable volume estimation}
Replacing the threshold acceptance-criterion by a probabilistic Metropolis criterion, and setting $N_\text{m}=1$, turns GaA into an adaptive MCMC sampler  with global adaptive scaling \cite{Andrieu:2008}. We termed this method {\em Metropolis-GaA} \cite{Muller:2010a,Muller:2010b}. Its strength is that GaA can automatically adapt to the covariance of the target probability distribution while maintaining the fixed hitting probability. For standard MCMC, this cannot be achieved without fine-tuning the proposal using multiple MCMC runs. We hypothesize that GaA might also be an effective tool for approximate Bayesian computation (ABC) \cite{Toni:2009}. In essence, the ABC ansatz is MCMC without an explicit likelihood function \cite{Marjoram:2003}. The likelihood is replaced by a distance function --- which plays the same role as our objective function --- that measures closeness between a parameterized model simulation and empirical data $\mathcal{D}$, or summary statistics thereof. When a uniform prior over the parameters and a symmetric proposal are assumed, a parameter vector in ABC is unconditionally accepted if its corresponding distance function value $f(\vv{\theta}^{(g+1)}) < c_\text{T}$ \cite{Marjoram:2003}. The threshold $c_\text{T}$ is a problem-dependent constant that is fixed prior to the actual computation. Marjoram and co-workers have shown that samples obtained in this manner are approximately drawn from the posterior parameter distribution given the data $\mathcal{D}$. 
While Pritchard \textit{et al.} used a simple rejection sampler \cite{Pritchard:1999}, Marjoram and co-workers proposed a standard MCMC scheme \cite{Marjoram:2003}. Toni and co-workers used sequential MC for sample generation \cite{Toni:2009}. To the best of our knowledge, however, the present work presents the first application of an adaptive MCMC scheme for ABC in biochemical network parameter inference. Finally, we emphasize that when GaA's mean, covariance matrix, and hitting probability $p$ stabilize during ABC, they provide direct access to an ellipsoidal estimation of the volume of the viable parameter space as defined by the threshold $c_\text{T}$ \cite{Muller:2011c}. Hafner and co-workers have shown how to use such viable volume estimates for model discrimination \cite{Hafner:2009}. 

\section{Evaluation of the forward model}
\label{sec:forwardmodel}

In each iteration of the GaA algorithm, the forward model of the network needs to be evaluated for the proposed parameter vector ${\vv{\theta}}$. This requires an efficient and exact SSA for the chemical kinetics of the reaction network, used to generate trajectories ${\vv{n}}(t)$ from $\mathcal{M}(\vv{\theta})$. Since GaA could well propose parameter vectors that lead to low copy numbers for some species, it is important that the SSA be exact since approximate algorithms are not appropriate at low copy number.    

In its original formulation, Gillespie's SSA has a computational cost that is linearly proportional to the total number $M$ of reactions in the network. If many model evaluations are required, as in the present application, this computational cost quickly becomes prohibitive. While more efficient formulations of SSA have been developed for weakly coupled reaction networks, their computational cost remains proportional to $M$ for strongly coupled reaction networks \cite{Ramaswamy:2009}. A reaction network is weakly coupled if the number of reactions that are influenced by any other reaction is bounded by a constant. If a network contains at least one reaction whose firing influences the propensities of a fixed proportion (in the worst case all) of the other reactions, then the network is strongly coupled \cite{Ramaswamy:2009}. Scale-free networks as seem to be characteristic for systems biology models \cite{Albert:2005,Strogatz:2001} are by definition strongly coupled. This is due to the existence of {\em hubs} that have a higher connection probability than other nodes. These hubs frequently correspond to chemical reactions that produce or consume species that also participate in the majority of the other reactions, such as water, ATP, or CO$_2$ in metabolic networks. 

We use partial-propensity methods \cite{Ramaswamy:2009,Ramaswamy:2010a} to simulate trajectories according to the solution of the chemical master Eq.~\ref{eq:CME}
of the forward model. Partial-propensity methods are exact SSAs whose computational cost scales at most linearly with the number $N$ of species in the network \cite{Ramaswamy:2009}. For large networks, this number is usually much smaller than the number of reactions. Depending on the network model at hand, different partial-propensity methods are available for its efficient simulation. Strongly coupled networks where the rate constants span only a limited spectrum of values are best simulated with the partial-propensity direct method (PDM) \cite{Ramaswamy:2009}. Multi-scale networks where the rate constants span many orders of magnitude are most efficiently simulated using the sorting partial-propensity direct method (SPDM) \cite{Ramaswamy:2009}. Weakly coupled reaction networks can be simulated at constant computational cost using the partial-propensity SSA with composition-rejection sampling (PSSA-CR) \cite{Ramaswamy:2010}. Lastly, reaction networks that include time delays can be exactly simulated using the delay partial-propensity direct method (dPDM) \cite{Ramaswamy:2011}. Different combinations of the algorithmic modules of partial-propensity methods can be used to constitute all members of this family of SSAs~\cite{Ramaswamy:2010a}. We refer to the original publications for algorithmic details, benchmarks of the computational cost, and a proof of exactness of partial-propensity methods.

\section{Objective function}
\label{sec:costfunction}
In the context of parameter identification of stochastic biochemical networks, a number of distance or objective functions have previously been suggested. Reinker \textit{et al.} proposed an approximate maximum-likelihood measure under the assumption that only a small number of reactions fire between two experimental measurement points, and a likelihood based on singular value decomposition that works when many reactions occur per time interval \cite{Reinker:2006}. Koutroumpas \textit{et al.} compared objective functions based on least squares, normalized cross-correlations, and conditional probabilities using a Genetic Algorithm \cite{Koutroumpas:2008}. Koeppl and co-workers proposed the Kantorovich distance to compare experimental and model-based probability distributions \cite{Koeppl:2010}. Alternative distance measures include the Earth Mover's distance or the Kolomogorov-Smirnov distance \cite{Poovathingal:2010}. These distance measures, however, can only be used when many experimental trajectories are available. In order to measure the distance between a {\em single} experimental trajectory $\hat{\vv{n}}(t)$ and a {\em single} model output $\vv{n}(t)$, we propose a novel cost function $f(\vv{\theta}) = f(\mathcal{M}(\vv{\theta}), \, \vv{\hat{n}})$ that reasonably captures the kinetics of a monostable system. We define a compound objective function $f(\vv{\theta}) = f_1(\vv{\theta}) + f_2(\vv{\theta})$ with
\begin{equation}
f_1(\vv{\theta}) = \sum_{i=1}^{4} \gamma_i\,\,, \qquad f_2(\vv{\theta}) = \sum_{i=1}^{N}\frac{\sum_{l=0}^{z_x}|\r{ACF}_l(\hat{n}_i) - \r{ACF}_l(n_i)|}{\sum_{l=0}^{z_x}\r{ACF}_l(\hat{n}_i)} \, ,   
\end{equation}
where
\begin{equation}
\gamma_i = \sum_{j=1}^{N} \sqrt{\left(\frac{\mu_{i}(n_j) - \mu_{i}(\hat{n}_j)}{\mu_{i}(\hat{n}_j)}\right)^2}  
\end{equation}
with the central moments given by
\begin{equation}
\mu_i(n_j) =\left\{
\begin{array}{l l}
    \sum_{p=1}^{K}n_j(t_0 + (p-1)\Delta t_\r{exp}) & \text{if} \quad i = 1\\
   ( | \sum_{q=1}^{K} \left(n_j(t_0 + (q-1)\Delta t_\r{exp}) - \mu_1(n_j)\right)^i | )^{1/i} & \, \text{otherwise}\\
 \end{array} 
   \right .
   \end{equation}
and the time-autocorrelation function (ACF) at lag $l$ given by
\begin{eqnarray}
\notag
\r{ACF}_l(n_i) = \frac{n_i(t_0) n_i(t_0+l\,\Delta t_\r{exp}) - (\mu_1(n_i))^2}{\mu_2(n_i)} \, .
\end{eqnarray}
The variable $z_x$ is the lag at which the experimental ACF crosses 0 for the first time. The function $f_1(\vv{\theta})$ measures the difference between the first four moments of $\vv{n}$ and $\vv{\hat{n}}$. This function alone would, however, not be enough to capture the kinetics since it lacks information about correlations in time. This is taken into account by $f_2(\vv{\theta})$, measuring the difference in the lifetimes of all chemical species. These lifetimes are systematically modulated by the volume $\Omega$~\cite{Ramaswamy:2011a}, hence enabling volumetric measurements of intra-cellular reaction compartments along with the identification of the rate constants. 

The present objective function allows inclusion of experimental readouts from image-based systems biology. The moment-matching part is a typical readout from fluorescence photometry, whereas the autocorrelation of the fluctuations can directly be measured using, e.g., FCS.       

\begin{figure}[b]
\sidecaption
\includegraphics[width=\columnwidth]{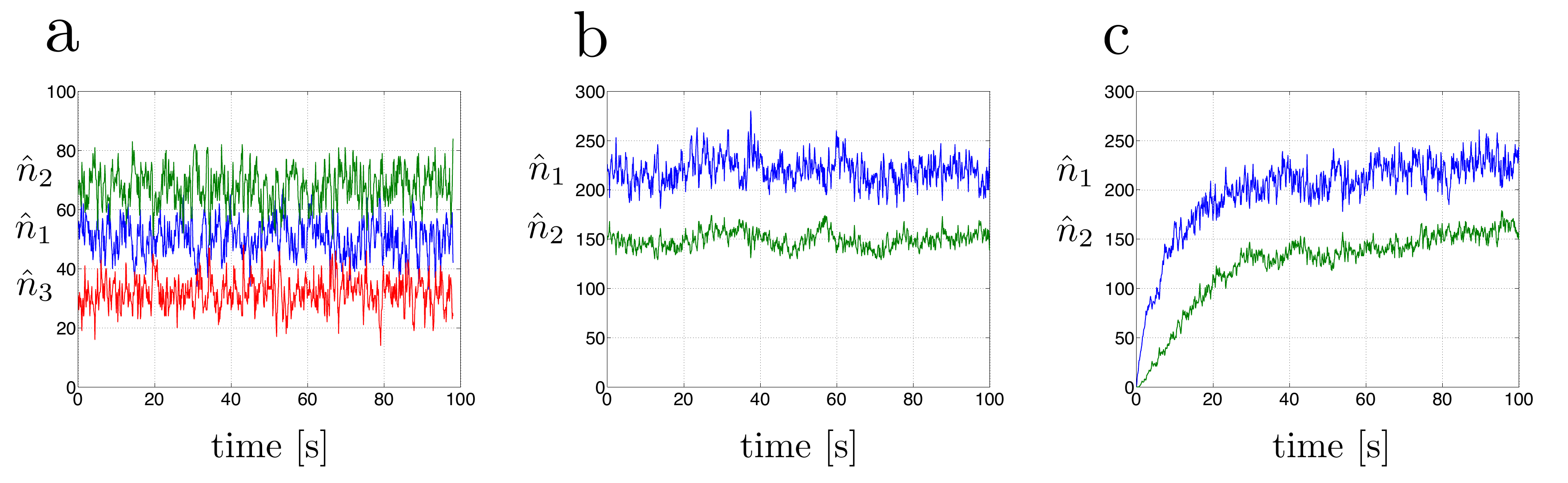}
\caption{\textit{In silico} data for all test cases. \textbf{a.} Time evolution of the populations of three species in the cyclic chain model at steady state (starting at $t_0=2000$). \textbf{b.} Time evolution of the populations of two species in the aggregation model at steady state (starting at $t_0=5000$). \textbf{c.} Same as  \textbf{b}, but during the transient phase (starting at $t_0=0$). }
\label{fig:trajData}      
\end{figure}

\section{Results}\label{sec:results}

We estimate the unknown parameters ${\vv{\theta}}$ for two reaction networks: a weakly coupled cyclic chain and a strongly coupled non-linear colloidal aggregation network. For the cyclic chain we estimate $\vv{\theta}$ at steady state. For the aggregation model we estimate $\vv{\theta}$ both at steady state and in the transient phase. Every kinetic parameter is allowed to vary in the interval $[10^{-3},\,10^3]$ and the reaction volume $\Omega$ in $[1, \, 500]$. Each GaA run starts from a point selected uniformly at random in logarithmic parameter space. 

\subsection{Weakly coupled reaction network: cyclic chain}
The cyclic chain network is given by:
\begin{eqnarray}
\notag
\r{S}_{i} \xrightarrow{k_{i}} & \r{S}_{i+1} &\qquad  i =1,\ldots ,N-1 \, ,\\
\r{S}_{i} \xrightarrow{k_{N}} & \r{S}_{1} &\qquad i =N \, . 
\label{eq:clc}
\end{eqnarray}
In this linear network, the number of reactions $M$ is equal to the number of species $N$. The maximum degree of coupling of this reaction network is 2,  irrespective of the size of the system (length of the chain), rendering it weakly coupled~\cite{Ramaswamy:2009}. We hence use PSSA-CR to evaluate the forward model with a computational complexity in $O(1)$ \cite{Ramaswamy:2010}. In the present test case, we limit ourselves to 3 species and 3 reactions, i.e., $N=M=3$. The parameter vector for this case is given by $\vv{\theta} = [k_1, k_2, k_3]$, since the kinetics of linear reactions is independent of the volume $\Omega$~\cite{Ramaswamy:2011a}.

We simulate steady-state ``experimental'' data ${\hat{\vv{n}}}$ using PSSA-CR with ground truth $k_1 = 2$, $k_2 = 1.5$, $k_3=3.2$ (see Fig.~\ref{fig:trajData}\textbf{a}). We set the initial population of the species to $n_1(t=0) = 50$, $n_2(t=0) = 50$, and $n_3(t=0) = 50$ and sample a single CME trajectory at equi-spaced time points with $\Delta t_\r{exp} = 0.1$ between  $t=t_0$ and $t=t_0+(K-1)\Delta t_\r{exp}$ with $t_0 = 2000$ and $K = 1001$ for each of the 3 species $\mathrm{S}_1$, $\mathrm{S}_2$, and $\mathrm{S}_3$. For the generated data we find $z_x=7$. 

We generate trajectories from the forward model for every parameter vector ${\vv{\theta}}$ proposed by GaA using PSSA-CR between $t=0$ and $t=(K-1)\Delta t_\r{exp} = 100$, starting from the initial population $n_i(t=0) = \hat{n}_i(t=t_0)$.

\begin{figure}[b]
\sidecaption
\includegraphics[width=\columnwidth]{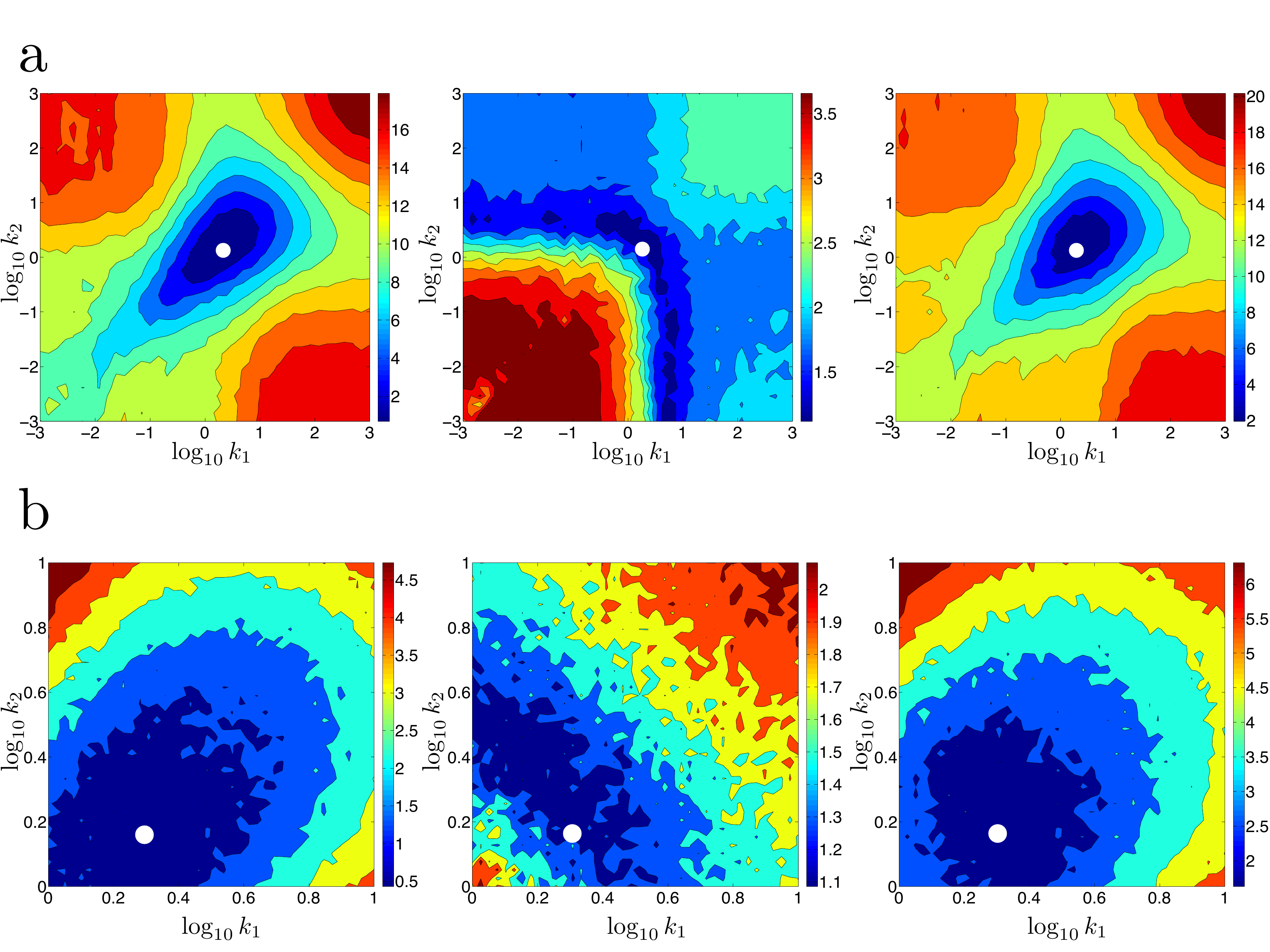}
\caption{\textbf{a.} Global objective function landscape for the cyclic chain over the complete search domain for optimal $k_3=3.2$. The three panels from left to right show $f_1(\vv{\theta})$, $f_2(\vv{\theta})$, and $f(\vv{\theta})$, respectively. \textbf{b.} A refined view of the global objective function landscape near the global optimum. The three panels from left to right show $f_1(\vv{\theta})$, $f_2(\vv{\theta})$, and $f(\vv{\theta})$, respectively. The white dots mark the true, optimal parameters.}
\label{fig:landscape}     
\end{figure}

Before turning to the actual parameter identification, we illustrate the topography of the objective function landscape for the present example. We fix $k_3=3.2$ to its optimal value and perform a two-dimensional grid sampling for $k_1$ and $k_2$ over the full search domain. We use 40 logarithmically spaced sample points per parameter, resulting in $40^2$ parameter combinations. For each combination we evaluate the objective function. The resulting landscapes of $f_1(\vv{\theta})$, $f_2(\vv{\theta})$, and $f(\vv{\theta})$ are depicted in Fig.~\ref{fig:landscape}\textbf{a}. Figure~\ref{fig:landscape}\textbf{b} shows refined versions around the global optimum.  
We see that the moment-matching term $f_1(\vv{\theta})$ is largely responsible for the global single-funnel topology of the landscape. The autocorrelation term $f_2(\vv{\theta})$ sharpens the objective function near the global optimum and renders it locally more isotropic.   

\begin{figure}[h]
\sidecaption
\includegraphics[width=\columnwidth]{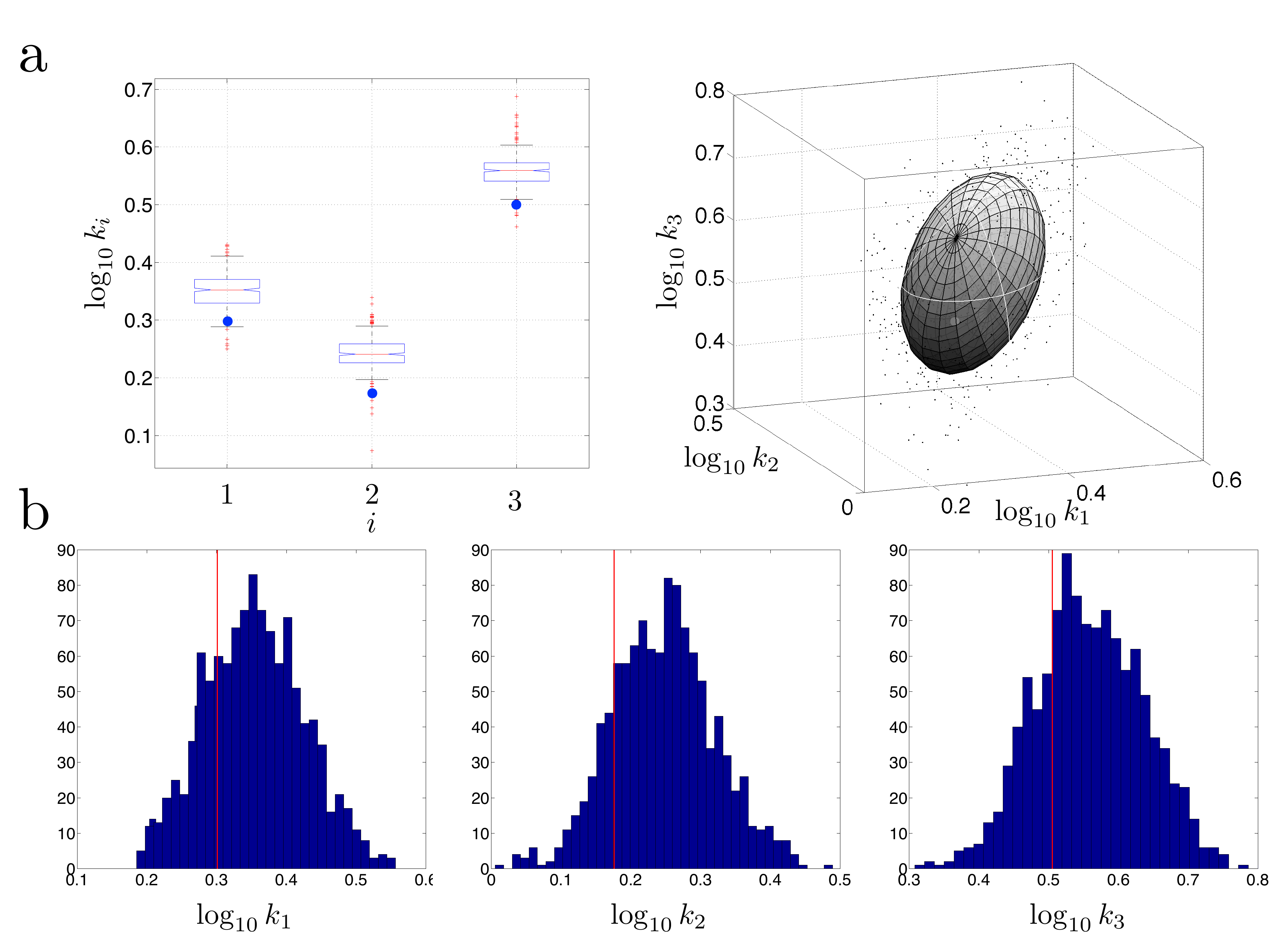}
\caption{\textbf{a.} Left panel: Box plot of the 30 best parameter vectors from each of the 15 independent optimization runs. The blue dots mark the true parameter values. Right panel: Ellipsoidal volume estimation of the parameter space below an objective-function threshold $c_\text{T}=2$ from a single ABC run. \textbf{b.} Empirical posterior distributions of the kinetic parameters from the same single ABC run with $c_\text{T}=2$. The red lines indicate the true parameters.}      
\label{fig:resCC}      
\end{figure}

We perform both global optimization and ABC runs using GaA. In each of the 15 independent optimization runs, the number of function evaluations (FES) is limited to MAX$\_$FES$=1000M = 3000$. We set the initial step size to $r^{(0)}=1$ and perform all searches in logarithmic scale of the parameters. Independent restarts from uniformly random points are performed when the step size $r$ drops below $10^{-4}$ \cite{Muller:2010}. For each of the 15 independent runs, the 30 parameter vectors with the smallest objective function value are collected and displayed in the box plot shown in the left panel of Fig.~\ref{fig:resCC}\textbf{a}. All 450 collected parameter vectors have objective function values smaller than 1.6. The resulting data suggest that the present method is able to accurately determine the correct scale of the kinetic parameters from a single experimental trajectory, although an overestimation of the rates is apparent. 

We use the obtained optimization results for subsequent ABC runs. We conduct 15 independent ABC runs using $c_\text{T}=2$. The starting points for the ABC runs are selected uniformly at random from the 450 collected parameter vectors in order to ensure stable initialization.     
For each run, we again set MAX$\_$FES$=1000M= 3000$. The initial step size $r^{(0)}$ is set to $0.1$, and the parameters are again explored in logarithmic scale. For all runs we observe rapid convergence of the empirical hitting probability $p_\text{emp}$ to the optimal $p=\frac{1}{\r{e}}$ (see Sec.~\ref{sec:GaA}). 
We collect the ABC samples along with the means and covariances of GaA as soon as $|p_\text{emp}-p| <0.05$. As an example we show the histograms of the posterior samples for a randomly selected run in Fig.~\ref{fig:resCC}\textbf{b}. The means of the posterior distributions are again larger than the true kinetic parameters. Using GaA's means, covariance matrices, and the corresponding hitting probabilities that generated the posterior samples, we can construct an ellipsoidal volume estimation \cite{Muller:2011c}.
This is done by multiplying each eigenvalue of the average of the collected covariance matrices with $c_{p_\text{emp}} = \text{inv}\,{\chi^{2}_{n}(p_\text{emp})}$, the $n$-dimensional inverse Chi-square distribution evaluated at the empirical hitting probability. The product of these scaled eigenvalues and the volume of the $n$-dimensional unit sphere, $|S(n)| = \frac{\pi^{\frac{n}{2}}}{\Gamma(\frac{n}{2}+1)}$, then yields the ellipsoid volume with respect to a uniform distribution (see \cite{Muller:2011c} for details).
The resulting ellipsoid contains the optimal kinetic parameter vector and is depicted in the right panel of Fig.~\ref{fig:resCC}\textbf{a}. It has a volume of $0.045$ in log-parameter space. This constitutes only 0.0208\% of the initial search space volume, indicating that GaA significantly narrows down the viable parameter space around the true optimal parameters. 

\subsection{Strongly coupled reaction network: colloidal aggregation}
The colloidal aggregation network is given by:
\begin{eqnarray}
\notag
\emptyset \xrightarrow{k_1^\r{on}} & \r{S}_1 \qquad &
\\
\notag
\r{S}_i + \r{S}_j \xrightarrow{k_{ij}} & \r{S}_{i+j} 
\qquad & 
i+j = 1, \ldots, N 
\\
\notag
\r{S}_{i+j} \xrightarrow{\bar{k}_{ij}} & \r{S}_i + \r{S}_j 
\qquad & 
i+j = 1, \ldots, N 
\\
\r{S}_i \xrightarrow{k_i^\r{off}} & \emptyset \qquad & i = 1, \ldots, N \, . 
\label{eq:agg}
\end{eqnarray}
For this network of $N$ species, the number of reactions is $M = \left\lfloor \frac{N^2}{2} \right\rfloor + N + 1$. The maximum degree of coupling of this reaction network is proportional to $N$, rendering the network strongly coupled~\cite{Ramaswamy:2009}. We hence use SPDM to evaluate the forward model with a computational complexity in $O(N)$ \cite{Ramaswamy:2009}. We use SPDM instead of PDM since the search path of GaA is unpredictable and could well generate parameters that lead to multi-scale networks. For this test case, we limit ourselves to two species, i.e., $N=2$ and $M = 5$. The parameter vector for this case is $\vv{\theta} = [k_{11}, \bar{k}_{11}, k_1^\r{on}, k_1^\r{off}, k_2^\r{off}, \Omega]$. 

We perform GaA global optimization runs following the same protocol as for the cyclic chain network with MAX\_FES = $1000(M+1) = 6000$.

\subsubsection{At steady state}
We simulate ``experimental'' data ${\hat{\vv{n}}}$ using SPDM with ground truth $k_{11} = 0.1$, $\bar{k}_{11} = 1.0$, $k_1^\r{on} = 2.1$, $k_1^\r{off} = 0.01$, $k_2^\r{off} = 0.1$, and $\Omega = 15$ (see Fig.~\ref{fig:trajData}\textbf{b}). We set the initial population of the species to $n_1(t=0) = 0$, $n_2(t=0) = 0$, and $n_3(t=0) = 0$ and sample K = 1001 equi-spaced data points between $t=t_0$ and $t=t_0+(K-1)\Delta t_\r{exp}$ with $t_0 = 5000$ and $\Delta t_\r{exp} = 0.1$. 

We generate trajectories from the forward model for every parameter vector ${\vv{\theta}}$ proposed by GaA using SPDM between $t=0$ and $t=(K-1)\Delta t_\r{exp} = 100$, stating from the initial population $n_i(t=0) = \hat{n}_i(t=t_0)$.

The optimization results are summarized in the left panel of Fig.~\ref{fig:resAgg}\textbf{a}.
For each of 15 independent runs, the 30 lowest-objective parameters are collected and shown in the box plot. We observe that the true parameters corresponding to $\theta_2 =  \bar{k}_{11}$, $\theta_3 =  k_1^\r{on}$, $\theta_4=k_1^\r{off}$, and $\theta_5 = k_2^\r{off}$ are between the $25^\r{th}$ and $75^\r{th}$ percentiles of the identified parameters. Both the first parameter and the reaction volume are, on average, overestimated. Upon rescaling the kinetic rate constants with the estimated volume, we find $\vv{\theta}^\text{norm} = [ \theta_1/\theta_6, \theta_2, \theta_3 \, \theta_6, \theta_4, \theta_5 ]$, which are the specific probability rates of the reactions. The identified values are shown in the right panel of Fig.~\ref{fig:resAgg}\textbf{a}. The median of the identified $\theta_3^\text{norm}$  coincides with the true specific probability rate. Likewise, $\theta_1^\text{norm}$ is closer to the $25^\r{th}$ percentile of the parameter distribution. This suggests a better estimation performance of GaA in the space of specific probability rates, at the expense of not obtaining an estimate of the reaction volume.

\subsubsection{In the transient phase}
We simulate ``experimental'' data in the transient phase of the network dynamics using the same parameters as above between $t=t_0$ and $t=(K-1)\Delta t_\r{exp}$ with $t_0 = 0$, $\Delta t_\r{exp} = 0.1$, and $K=1001$ (see Fig.~\ref{fig:trajData}\textbf{c}). We evaluate the forward model with $n_i(t=0) = \hat{n}_i(t=t_0)$ to obtain trajectories between $t=0$ and $t=(K-1)\Delta t_\r{exp}$ for every proposed parameter vector ${\vv{\theta}}$.    

\begin{figure}[b]
\sidecaption
\includegraphics[width=\columnwidth]{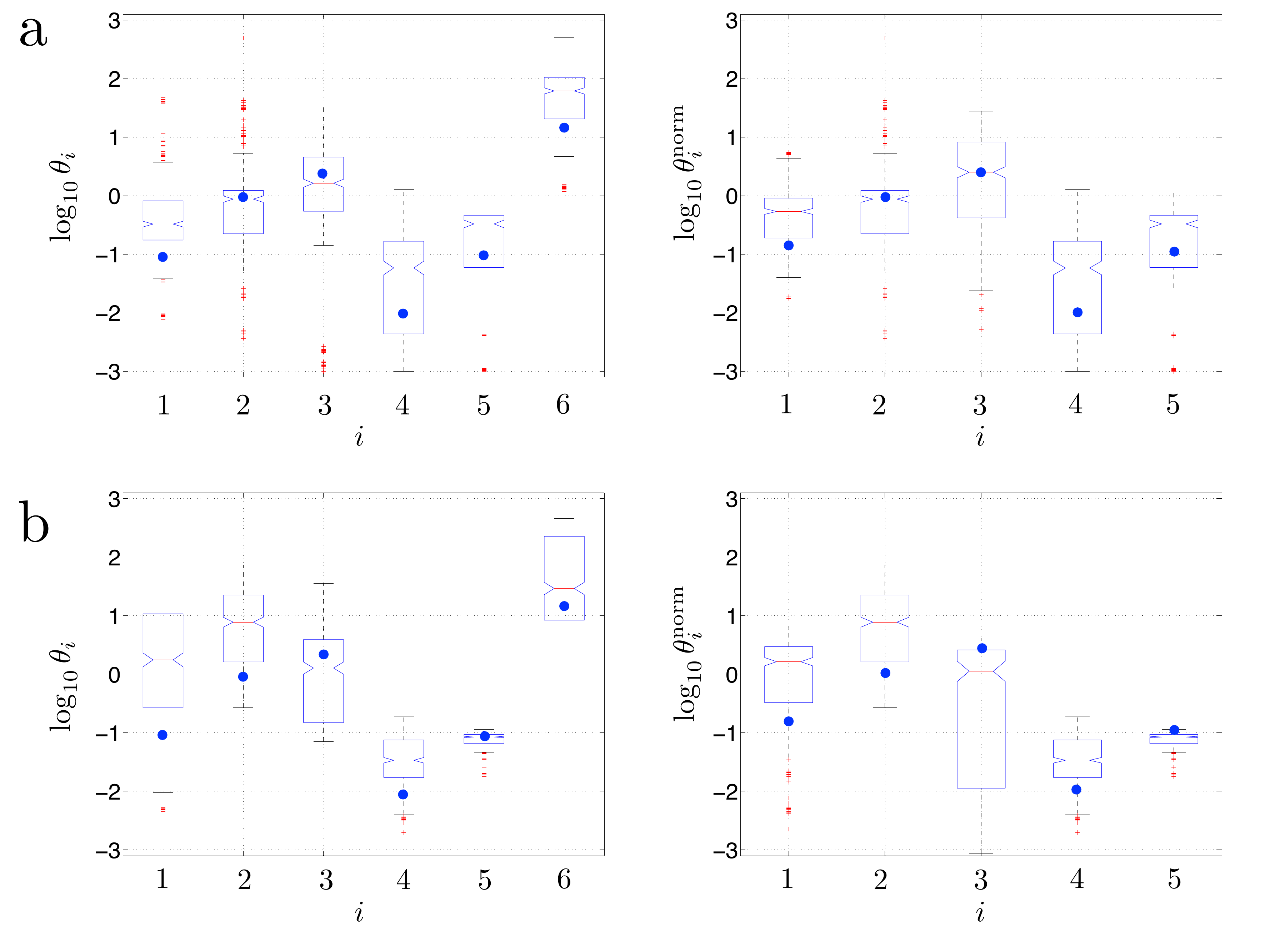}
\caption{\textbf{a.} Left panel: Box plot of the 30 best parameter vectors from each of the 15 independent optimization runs for the steady-state data set. Right panel: Box plots of the normalized parameters (see main text for details). \textbf{b.} Left panel: Box plot of the 30 best parameter vectors from each of the 15 independent optimization runs for the transient data set. Right panel: Box plot of the normalized parameters (see main text for details). The blue dots indicate the true parameter values.}      
\label{fig:resAgg}
\end{figure}

The optimization results for the transient case are summarized in Fig.~\ref{fig:resAgg}\textbf{b}.
We observe that the true parameters corresponding to $\theta_3 =  k_1^\r{on}$, $\theta_5 = k_2^\r{off}$, and $\theta_6 =  \Omega$ are between the $25^\r{th}$ and $75^\r{th}$ percentiles of the identified parameters. The remaining parameters are, on average, overestimated. In the space of rescaled parameters $\vv{\theta}^\text{norm}$ we do not observe a significant improvement of the estimation. \\

\section{Conclusions and Discussion}\label{sec:discuss}
We have considered parameter estimation of stochastic biochemical networks from single experimental trajectories. Parameter identification from single time series is desirable in image-based systems biology, where per-cell estimates of the fluorescence evolution and its fluctuations are available. This enables quantifying cell--cell variability on the level of network parameters. The histogram of the parameters identified for different cells provides a biologically meaningful way of assessing phenotypic variability beyond simple differences in the fluorescence levels.

We have proposed a novel combination of a flexible Monte Carlo method, the Gaussian Adaptation (GaA) algorithm, and efficient exact stochastic simulation algorithms, the partial-propensity methods. The presented method can be used for global parameter optimization, approximate Bayesian inference under a uniform prior, and volume estimation of the viable parameter space. We have introduced an objective function that measures closeness between a single experimental trajectory and a single trajectory generated by the forward model. The objective function comprises a moment-matching and a time-autocorrelation part. This allows including experimental readouts from, e.g., fluorescence photometry and fluorescence correlation spectroscopy. 

We have applied the method to estimate the parameters of two monostable reaction networks from a single simulated temporal trajectory each, both at steady state and during transient phases. We considered the linear cyclic chain network and a non-linear colloidal aggregation network. For the linear model we were able to 
robustly identify a small region of parameter space containing the true kinetic parameters. In the non-linear aggregation model, we could identify several parameter vectors that fit the simulated experimental data well. There are two possible reasons for this reduced parameter identifiability: either GaA cannot find the globally optimal region of parameter space due to high ruggedness and noise in the objective function, or the non-linearity of the aggregation network modulates the kinetics in a non-trivial way \cite{Ramaswamy:2011a,Grima:2009}. Both cases are not accounted for in the current objective function, thus leading to reduced performance for non-linear reaction networks.

We also used GaA as an adaptive MCMC method for approximate Bayesian inference of the posterior parameter distributions in the linear chain network. This enabled estimating the volume of the viable parameter space below a given objective-function value threshold. We found these volume estimates to be stable across independent  runs. We thus believe that GaA might be a useful tool for exploring the parameter spaces of stochastic systems. 

Future work will include (i) alternative objective functions that include temporal cross-correlations between species and the derivative of the autocorrelation; (ii) longer experimental trajectories; (iii) multi-stable and oscillatory systems; and (iv) alternative global optimization schemes.   
Moreover, the applicability of the present method to large-scale, non-linear biochemical networks and real-world experimental data will be tested in future work.  

\begin{acknowledgement}
RR was financed by a grant from the Swiss SystemsX.ch initiative (grant WingX), evaluated by the Swiss National Science Foundation. This project was also supported with a grant from the Swiss SystemsX.ch initiative, grant LipidX-2008/011, to IFS.
\end{acknowledgement}

\end{document}